\documentclass[10pt,onecolumn,draftcls,journal]{IEEEtran}
\usepackage[ansinew]{inputenc}
\usepackage[T1]{fontenc}
\usepackage{color,amsmath,enumerate,amssymb,hhline,verbatim}
\usepackage{url}
\usepackage{multirow}

\DeclareMathOperator{\rank}{rank}
\begin{document}

\title{Constructions and Properties of Linear Locally Repairable Codes
\thanks{Part of this work was presented at Global Wireless Summit 2014, Aalborg, Denmark \cite{LrcGws}. Also, part of the paper is to be presented at IEEE Information Theory Workshop 2014.}
}

\author{Toni Ernvall, Thomas Westerb{\"a}ck,  Camilla Hollanti and Ragnar Freij
\thanks{T. Ernvall is with Turku Centre for Computer Science, Turku, Finland and with the Department of Mathematics and Statistics, University of Turku, Finland (e-mail:tmernv@utu.fi).}

\thanks{T. Westerb{\"a}ck and C. Hollanti are with the Department of Mathematics and Systems Analysis, Aalto University (e-mails:thomas.westerback@aalto.fi, camilla.hollanti@aalto.fi).}

\thanks{R. Freij is with the Department of Communications and Networking, Aalto University (e-mail:ragnar.freij@aalto.fi).}

\thanks{T. Westerbäck and C. Hollanti are financially supported by the Academy of Finland grants \#276031, \#282938, and \#283262, and by a grant from Magnus Ehrnrooth Foundation, Finland. The support from the European Science Foundation under the ESF COST Action IC1104 is gratefully acknowledged.}
}

\maketitle

\newtheorem{definition}{Definition}[section]
\newtheorem{thm}{Theorem}[section]
\newtheorem{proposition}[thm]{Proposition}
\newtheorem{lemma}[thm]{Lemma}
\newtheorem{corollary}[thm]{Corollary}
\newtheorem{exam}{Example}[section]
\newtheorem{conj}{Conjecture}
\newtheorem{remark}{Remark}[section]

\newcommand{\La}{\mathbf{L}}
\newcommand{\h}{{\mathbf h}}
\newcommand{\Z}{{\mathbf Z}}
\newcommand{\R}{{\mathbf R}}
\newcommand{\C}{{\mathbf C}}
\newcommand{\D}{{\mathcal D}}
\newcommand{\F}{{\mathbf F}}
\newcommand{\HH}{{\mathbf H}}
\newcommand{\OO}{{\mathcal O}}
\newcommand{\G}{{\mathcal G}}
\newcommand{\A}{{\mathcal A}}
\newcommand{\B}{{\mathcal B}}
\newcommand{\I}{{\mathcal I}}
\newcommand{\E}{{\mathcal E}}
\newcommand{\PP}{{\mathcal P}}
\newcommand{\Q}{{\mathbf Q}}
\newcommand{\M}{{\mathcal M}}
\newcommand{\separ}{\,\vert\,}
\newcommand{\abs}[1]{\vert #1 \vert}

\begin{abstract}
In this paper, locally repairable codes with all-symbol locality are studied. Methods to modify already existing codes are presented. Also, it is shown that with high probability, a random matrix with a few extra columns guaranteeing the locality property, is a generator matrix for a locally repairable code with a good minimum distance. The proof of this also gives a constructive method to find locally repairable codes. Constructions are given of three infinite classes of optimal vector-linear locally repairable codes over an alphabet of small size, not depending on the size of the code.
\end{abstract}


%
\IEEEpeerreviewmaketitle

\section{Introduction}
\subsection{Locally Repairable Codes}
In the literature, three kinds of repair cost metrics are studied: \emph{repair bandwidth} \cite{dimakis}, \emph{disk-I/O} \cite{diskIO}, and \emph{repair locality} \cite{Gopalan,Oggier,Simple}. In this paper, the repair locality is the subject of interest.

Given a finite set $\mathbb{A}$, and an injective function $f: \mathbb{A}^k \rightarrow \mathbb{A}^n$, let $C$ denote the image of $f$.
We say that $C$ is a \emph{locally repairable code (LRC)} and has \emph{all-symbol $(r,\delta)$-locality} with parameters $(n,k,d)$, if the code $C$ has minimum (Hamming) distance $d$ and all the $n$ symbols of the code have $(r,\delta)$-locality. The concept was introduced in \cite{prakash}. An $(r,\delta)$-locality for the $j$th symbol is defined to be a subset $S_j \subseteq \{1,\dots,n\}$ such that $j \in S_j$, $|S_j| \leq r+\delta-1$, and the code restricted to code symbols in $S_j$ has minimum distance at least $\delta$. In particular, the $j$th symbol in a code word is determined by any choice of $|S_j|-\delta+1$ symbols from $S_j$.
LRCs are defined when $1 \leq r \leq k$ and $\delta \geq 2$. By a linear LRC we mean that the code is a $k$-dimensional subspace of $\mathbb{F}_{q}^{n}$, where $\mathbb{F}_q$ is the finite field with $q$ elements.

In \cite{prakash} it is shown that we have the following bound for a linear locally repairable code $C$ of length $n$, dimension $k$, minimum distance $d$ and all-symbol $(r,\delta)$-locality:
\begin{equation}\label{upperbound}
d \leq n - k - \left( \left \lceil \frac{k}{r} \right \rceil -1 \right)\left( \delta - 1 \right) + 1.
\end{equation}
A linear LRC  that meets this bound is called \emph{optimal}. For this reason we write \[d_{\text{opt}}(n,k,r,\delta)=n - k - \left( \left \lceil \frac{k}{r} \right \rceil -1 \right)\left( \delta - 1 \right) + 1.\]

Linear LRCs are \emph{scalar} in the sense that each code symbols is an element of a finite field. In \cite{silberstein13} the concept of scalar LRCs was generalized to \emph{vector} LRCs, where each code symbol is a vector over some finite field. A \emph{vector-linear} code over $\mathbb{F}_q^m$ is a vector code which is linear over $\mathbb{F}_q$, with code alphabet $\mathbb{F}_q^m$. An LRC with $\delta = 2$ is called an $(n,k,d,r)$-LRC. Therefore, a \emph{vector-linear $(n,k,d,r)$}-LRC over $\mathbb{F}_q^m$ is a subset $C \subseteq (\mathbb{F}_q^m)^n$ so that $|C| = q^{mk}$, $C$ is a linear code over $\mathbb{F}_q$, the minimum distance is $d$ in the context of the code alphabet $\mathbb{F}_q^m$, and all the code symbols have $(r,2)$-locality, again in the context of the $\mathbb{F}_q^m$-alphabet. Note that a vector-linear $(n,k,d,r)$-LRC over $\mathbb{F}_q^m$ also can been seen as a (possibly non-linear) $(n,k,d,r)$-LRC over $\mathbb{F}_{q^m}$. A generalization of the bound given in \eqref{upperbound} for linear and non-linear codes was derived in \cite{LRCpapailiopoulos}. In our setting of vector-linear $(n,k,d,r)$-LRCs, this bound gives that
\begin{equation} \label{eq:upper_bound_r_locality}
d \leq n - k - \left \lceil \frac{k}{r}  \right \rceil + 2.
\end{equation}
The bound given above in \eqref{eq:upper_bound_r_locality} is also valid for both linear and non-linear $(n,k,d,r)$-LRCs. Therefore, a (linear, non-linear, vector-linear) LRC achieving the  bound in \eqref{eq:upper_bound_r_locality} is called \emph{optimal}.

\subsection{Related Work}
In \cite{LRCmatroid}, \cite{SongOptimal}, \cite{Rawat} and \cite{TamoBarg} the existence of optimal LRCs was proved for several values of the parameters $(n,k,r)$. Good codes, with the weaker assumption of information symbol locality, are designed in \cite{Pyramid}. In \cite{Gopalan} it was shown that there exist parameters $(n,k,r)$ for linear LRCs for which the bound of~\eqref{upperbound} is not achievable. LRCs corresponding to MSR and MBR points are studied in \cite{Kamath}.

Constructions of optimal $(n,k,d,r)$-LRCs over small finite fields were stated as an open problem for LRCs in~\cite{LRCmatroid}. Small finite fields as code alphabets are often desirable for practical reasons \cite{goparaju14}. A family of optimal linear $(n,k,d,r)$-LRCs over $\mathbb{F}_q$, generalizing the Reed-Solomon construction, is given in \cite{TamoBarg}, for any $q \geq n$. In \cite{goparaju14}, a construction is given of a class of optimal linear $(n,k,d,r)$-LRCs over $\mathbb{F}_2$. An upper bound similar to the bound given in \eqref{eq:upper_bound_r_locality}, taking the field size into account, is given in \cite{cadambe13}.

\subsection{Contributions and Organization}
In this paper, we will study codes with all-symbol locality, for given parameters $n$, $k$, $r$, and $\delta$. We will present methods to modify an already existing code to find smaller and larger codes. On some occasions, when the starting point is optimal, the resulting code is also optimal. We also show that a random matrix, with a few non-random extra columns to guarantee the repair property, generates a linear LRC with good minimum distance, with probability approaching one as the field size approaches infinity. It should be noted that all the results, except those considering small fields, are proven using only elementary results from linear algebra. However, we use the concept of circuits from matroid theory in the narrow sense where it has a simple interpretation in the language of linear algebra. All proofs in this paper are constructive.

Using a construction of quasi-uniform codes, given in \cite{thomas13}, we construct optimal vector-linear LRCs over $\mathbb{F}_2^2$ with parameters $(n,k,d,r)$ equal to $(4i+3,3i+1,3,3)$, $(4i+4,3i+2,3,3)$ and $(4i+4,3i+1,4,3)$ for $i \geq 1$.

Section \ref{Sec:BuildingCodes} gives two procedures to exploit already existing codes when building new ones. To be exact, it explains how we can build a new linear code of length $n+1$ and dimension $k+1$ with all-symbol $(r+1,\delta)$-locality from an already existing linear code of length $n$ and dimension $k$ with all-symbol $(r,\delta)$-locality, such that the minimum distance remains the same.

The same section also introduces a method to find a smaller code when given a code associated to parameters $(n,k,r,\delta)$. Namely, the procedure gives a code of length $n-1$, dimension $k-1$, minimum distance $d' \geq d$ and all-symbol $(r,\delta)$-locality.

In Section \ref{Sec:minDistance}, we give a construction of almost optimal linear locally repairable codes, with all-symbol $(r,\delta)$-locality. By almost optimal we mean that the minimum distance of a code is at least $d_{\text{opt}}(n,k,r,\delta)-\delta+1$.

In Section \ref{Sec:Random}, we study random matrices with a few non-random extra columns that guarantee the repair property. Using the construction of Section \ref{Sec:minDistance}, it is shown that these random codes perform well with high probability.

In Section \ref{sec:construction}, we give constructions of three classes of optimal vector-linear LRCs over $\mathbb{F}_2^2$. These constructions are based on a construction of quasi-uniform codes.

\section{Building Codes from Other Codes}\label{Sec:BuildingCodes}

\subsection{Some Technical Facts}
In this section, we will study how one can modify a locally repairable code to get a bigger or a smaller code, in terms of length. Strictly speaking, we will show how one can build a new linear code of length $n+1$ and dimension $k+1$ with all-symbol repair locality $(r+1,\delta)$, from a linear code of length $n$ and dimension $k$ with all-symbol repair locality $(r,\delta)$, such that the minimum distance remains the same. Also, we will show how to find a code for parameters \[(n'=n-1,k' = k-1,d' \geq d,r'=r).\]

Before stating the results, we need some definitions. Throughout this paper, $q$ is a prime power and $\mathbb{F}_q$ is a finite field with $q$ elements. Let $\mathbf{x},\mathbf{y} \in \mathbb{F}_q^n$. Then $d(\mathbf{x},\mathbf{y})$ is the Hamming distance of vectors $\mathbf{x}$ and $\mathbf{y}$. The weight of $\mathbf{x}$ is $w(\mathbf{x})=d(\mathbf{x},\mathbf{0})$.
The sphere with radius $s$ and center $\mathbf{x}$ is defined as
\[
B_s(\mathbf{x})=\{\mathbf{y} \in \mathbb{F}_q^n \mid d(\mathbf{x},\mathbf{y}) \leq s\}.
\]
The cardinality of the sphere is
\[
V_q(n,s)=|B_s(\mathbf{x})|= \sum_{i=0}^{s} \binom{n}{i} (q-1)^i,
\]
for which we have a trivial upper bound
\[
V_q(n,s) \leq (1+s) \binom{n}{\lfloor \frac{n}{2} \rfloor} q^s.
\]

We will also need a simple lemma.
\begin{lemma}\label{Lemma:lowerbound}
Let $n$ be a positive integer, and let $x$ and $c_j$ be nonnegative numbers, with $x \geq c_j$ for $j=1,\dots,n$. Then
\[
\prod_{j=1}^{n}(x-c_j) \geq x^n - \sum_{j=1}^{n}c_{j}x^{n-1}
\]
\end{lemma}
\begin{IEEEproof}
We will show this by induction. If $n=1$ the claim is clear. Assume the claim to be true for $n=m-1$ with $m \geq 2$.
Now
\begin{equation}
\begin{split}
\prod_{j=1}^{m}(x-c_j) & =(x-c_m)\prod_{j=1}^{m-1}(x-c_j) \\
& \geq (x-c_m)\left( x^{m-1} - \sum_{j=1}^{m-1}c_{j}x^{m-2} \right) \\
& = x^{m} - \sum_{j=1}^{m}c_{j}x^{m-1} + c_{m}\sum_{j=1}^{m-1}c_{j}x^{m-2} \\
& \geq x^{m} - \sum_{j=1}^{m}c_{j}x^{m-1}.
\end{split}
\end{equation}
\end{IEEEproof}

It is easy to verify that in a linear code generated by the matrix $(\mathbf{x}_1|\dots|\mathbf{x}_n)$ the $j$th node can be repaired using nodes $\mathbf{x}_{i_1},\dots,\mathbf{x}_{i_r}$ if and only if these vectors span a subspace to which $\mathbf{x}_j$ belongs. For this reason we adopt a definition of \emph{circuit} from matroid theory. For the connections between matroid theory and locally repairable codes, an interested reader is referred to \emph{e.g.} \cite{LRCmatroid}.

\begin{definition}
Consider a matrix $(\mathbf{x}_1|\dots|\mathbf{x}_n)$. A subset $\{i_1,\dots,i_s\} \subseteq \{1,\dots,n \}$ of size $s$ is called a circuit if $\{\mathbf{x}_{i_1},\dots,\mathbf{x}_{i_s}\}$ is linearly dependent, but all its proper subsets are linearly independent.
\end{definition}

It is easy to check that under the assumption of linear codes and all-symbol $(r,\delta)$-locality, for each index $j=1,\dots,n$ there must exist a subset \[\{i_1,\dots,i_{r+\delta-2}\} \subseteq \{1,\dots,n \} \setminus \{j\}\] such that any $r$ column matrices corresponding to $r$ elements of $\{i_1,\dots,i_{r+\delta-2}\}$ span a subspace to which the $j$th column vector belongs.

\subsection{Enlarging codes}
Now we will study how to enlarge codes. If $r=k$ then we always get an optimal linear LRC by \emph{a maximum distance separable} code, \emph{i.e.}, a linear code of dimension $n-d+1$ with $d$ being the minimum distance. Hence in this section we will assume that $r<k$.

As an example, let us first consider how to enlarge already existing codes in the case $\delta=2$.
\begin{exam}
Suppose we have a linear locally repairable code $C$ of length $n$, dimension $k$, minimum distance $d$, and $(r,2)$ all-symbol locality. Write
\[
\begin{pmatrix}
   a_{1,1} & \cdots & a_{1,n} \\
   \vdots  & \ddots & \vdots  \\
   a_{k,1} & \cdots & a_{k,n}
  \end{pmatrix}
\]
to be its generator matrix. Assume also that the code is built over a field of size $q > d \binom{n}{\left\lfloor \frac{n}{2} \right\rfloor}$. By Equation (\ref{upperbound}) we know that
\[
d+k-1 \leq n - \left \lceil \frac{k}{r} \right \rceil  + 1 \leq n -2 +1=n-1
\]
and hence
\begin{displaymath}
\begin{split}
|C|\cdot V_q(n,d-1) & \leq q^k \cdot (1+d-1) \binom{n}{\left\lfloor \frac{n}{2} \right\rfloor} q^{d-1}\\
& = d \binom{n}{\left\lfloor \frac{n}{2} \right\rfloor} q^{k+d-1} \\
& \leq d \binom{n}{\left\lfloor \frac{n}{2} \right\rfloor} q^{n-1}<q^n.
\end{split}
\end{displaymath}
Therefore there exists a vector $\mathbf{y} \in \mathbb{F}_{q}^{n}$ of distance at least $d$ to all the code vectors. Write $\mathbf{y}=(y_{1}, \dots,y_{n})^{t}$ and define two matrices $G_1$ and $G_2$ to be
\[
\begin{pmatrix}
   a_{1,1} & \cdots & a_{1,n} \\
   \vdots  & \ddots & \vdots  \\
   a_{k,1} & \cdots & a_{k,n} \\
   y_{1}   & \cdots & y_{n}
  \end{pmatrix}
\text{ and }
\begin{pmatrix}
   a_{1,1} & \cdots & a_{1,n} & 0 \\
   \vdots  & \ddots & \vdots  & \vdots  \\
   a_{k,1} & \cdots & a_{k,n} & 0 \\
   y_{1}   & \cdots & y_{n}   & 1
  \end{pmatrix},
\]
respectively. Here, the matrices have rows indexed by code words, and columns indexed by symbols of the codes.

Now, in the code generated by $G_2$, nodes $1,\dots,n$ have a locality of size at most $r+1$. If the $(n+1)$th node does not have a locality of size at most $r+1$ then in the code generated by $G_1$ all the nodes have a locality of size at most $r$. Hence, we either get a locally repairable code with all-symbol locality corresponding to the parameters $(n,k+1,r)$ or $(n+1,k+1,r+1)$. In both cases the minimum distance is still $d$. Indeed, let $\mathbf{u}=a\mathbf{y}+\mathbf{z} \neq \mathbf{0}$ where $a \in \mathbb{F}_q$ and $\mathbf{z} \in C$. Now if $a=0$ we have
\[
w(\mathbf{u})=w(\mathbf{z}) \geq d
\]
and if $a \neq 0$ we have
\[
w(\mathbf{u})=w(a^{-1}\mathbf{u})=w(\mathbf{y}+a^{-1}\mathbf{z})=d(\mathbf{y},-a^{-1}\mathbf{z}) \geq d
\]
proving the claim for the code generated by $G_1$. If we puncture the last symbol of the code generated by $G_2$ we get the code generated by $G_1$ and hence also its minimum distance is $d$.
\end{exam}

When $\delta >2$ the situation is slightly more complicated compared to the example. The next theorem gives the generalization for it.
\begin{thm}\label{Thm:Enlarge}
Suppose we have a linear LRC for parameters $(n,k,d,r,\delta)$ over a field $\mathbb{F}_q$, with \[q > (r+d) \binom{n}{\left\lfloor \frac{n}{2} \right\rfloor}\] and $r<k$. Then there exists a linear LRC for parameters \[(n'=n+1,k'=k+1,d'=d,r'=r+1,\delta'=\delta),\] over the same field.
\end{thm}
\begin{IEEEproof}
Let $C$ be a linear LRC for parameters $(n,k,d,r,\delta)$, over a field $\mathbb{F}_q$ with \[q >  (r+d) \binom{n}{\left\lfloor \frac{n}{2} \right\rfloor}.\] Let $G$ be its generator matrix, \emph{i.e.}, $G$ is a $k \times n$ matrix such that its row vectors form a basis for $C$. Write
\[
G=(\mathbf{x}_1|\dots|\mathbf{x}_n),
\]
where $\mathbf{x}_j \in \mathbb{F}_{q}^{k}$ for all $j=1,\dots,n$. Without loss of generality we may assume that $\mathbf{x}_j \neq \mathbf{0}$ for all $j=1,\dots,n$, since otherwise we could drop the zero columns off at this point, and at the end of the proof add the same number of zero columns into the maintained generator matrix.

Define a set $A \subseteq \mathbb{F}_{q}^{n}$, consisting of vectors \[(a_1,\dots , a_n)\in \mathbb{F}_{q}^{n}\] such that the following holds for every $j\in \{1,\dots,n\}$:
For every circuit $\{i_1,\dots , i_{s+1}\} \subseteq \{1,\dots,n\}$ of the matroid generated by $G$, with $i_1 \leq i_2 \leq \dots \leq i_{s+1}=j$ and $s\leq r$, consider a linear relation
\[
b_1\mathbf{x}_{i_1}+b_2\mathbf{x}_{i_2}+\dots+b_{s+1}\mathbf{x}_{i_{s+1}}=\mathbf{0}
\] between the symbols in the circuit.
Define $a_j$ to be any element of $\mathbb{F}_{q}$ such that
\[
b_1a_{i_1}+b_2a_{i_2}+\dots+b_{s+1}a_{i_{s+1}} \neq 0.
\]
For each $a_j$ ($j=1,\dots,n$) there are at least $q-c_j$ possibilities to choose from, where $c_j$ is the number of circuits with greatest element being $j$.

We have
\[
\sum_{j=1}^{n} c_j \leq \binom{n}{2}+\binom{n}{3}+\dots+\binom{n}{r+1} \leq r\binom{n}{\left\lfloor\frac{n}{2}\right\rfloor},
\]
and by using Lemma \ref{Lemma:lowerbound}, we see that the cardinality of $A$ is at least
\[
\prod_{j=1}^{n} (q - c_j) \geq q^n - r\binom{n}{\left\lfloor\frac{n}{2}\right\rfloor}q^{n-1}.
\]

Let $B \subseteq \mathbb{F}_{q}^{n}$ be the set of vectors with distance at least $d$ to the code vectors. Notice first that Equation (\ref{upperbound}) gives
\[\begin{split}
d & \leq n - k - \left( \left \lceil \frac{k}{r} \right \rceil -1 \right)\left( \delta - 1 \right) + 1\\
 & \leq n - k - \left( 2 -1 \right) + 1\\
 &=n-k.
\end{split}\]
Now
\[\begin{split}
|C|V_q(n,d-1) & \leq q^k \cdot (1+d-1) \binom{n}{\left\lfloor \frac{n}{2} \right\rfloor} q^{d-1}\\
& = d \binom{n}{\left\lfloor \frac{n}{2} \right\rfloor} q^{k+d-1} \\
& \leq d \binom{n}{\left\lfloor \frac{n}{2} \right\rfloor} q^{n-1},
\end{split}\]
and hence
\[\begin{split}
|B| &\geq q^n-|C|V_q(n,d-1) \\
& \geq q^n-d \binom{n}{\left\lfloor \frac{n}{2} \right\rfloor} q^{n-1}.
\end{split}\]

This implies that
\[\begin{split}
|A \cap B| =\, & |A|+|B|-|A \cup B| \\
\geq & \left( q^n - r\binom{n}{\left\lfloor\frac{n}{2}\right\rfloor}q^{n-1} \right) \\
& + \left( q^n-d \binom{n}{\left\lfloor \frac{n}{2} \right\rfloor} q^{n-1} \right) -q^n \\
=\, & q^n - (r+d)\binom{n}{\left\lfloor\frac{n}{2}\right\rfloor}q^{n-1} \\
>\, & 0,
\end{split}\]
and hence there exists a vector $\mathbf{a}$ in $A \cap B$.

Denote by $G_2$ a new $(k+1) \times (n+1)$ matrix
\[
\left(
\begin{array}{c|c}
G & \mathbf{0} \\ \hline
\mathbf{a}^t & 1
\end{array}
\right),
\]
where $\mathbf{0}$ is an all-zero vector from $\mathbb{F}_q^k$. Write also \[G_2=(\mathbf{y}_1|\dots|\mathbf{y}_{n+1}).\]

Denote by $C_2$ a code generated by $G_2$. Clearly $C_2 \subseteq \mathbb{F}_q^{n+1}$ and its dimension is $k+1$. Its minimum distance is~$d$: Let \[\mathbf{u}=a\mathbf{y}+\mathbf{z} \neq \mathbf{0},\] where $a \in \mathbb{F}_q$, $\mathbf{y}^t=(\mathbf{a}^t|1)$, and $\mathbf{z}^t=(\mathbf{z'}^t|0)$ with $\mathbf{z'}$ being a vector from $C$. Now if $a=0$ we have
\[
w(\mathbf{u})=w(\mathbf{z})=w(\mathbf{z'}) \geq d,
\]
and if $a \neq 0$ we have
\[\begin{split}
w(\mathbf{u})& =w(a^{-1}\mathbf{u})=w(\mathbf{y}+a^{-1}\mathbf{z})\\
& =d(\mathbf{y},-a^{-1}\mathbf{z})=d(\mathbf{a},-a^{-1}\mathbf{z'})+d(1,0) \\
& \geq d+1.
\end{split}\]

Let \[\mathbf{e}=(0,\dots,0,1)^t=\mathbf{y}_{n+1}\] be a column vector in $\mathbb{F}_{q}^{k+1}$. Write also \[\mathbf{y_j}^t=(\mathbf{x_j}^t|a_j)\] for $j=1,\dots,n$.

The code $C_2$ has $(r+1,\delta)$ repair locality for all symbols: Suppose $\{ i_1,\dots,i_{s+\delta-1} \}$ is an $(s,\delta)$-locality for the $i_1$th node in the original system. We will next show that $\{ i_1,\dots,i_{s+\delta-1}, n+1 \}$ is an $(s+1,\delta)$-locality for both the $i_1$th and $(n+1)$th node in the new system. First we will show that this is true for the $i_1$th node: Let \[S \subseteq \{ i_1,\dots,i_{s+\delta-1}, n+1 \}\setminus\{ i_1 \}\] be a subset with $|S|=s+1$. Write $S=\{j_1,\dots,j_{s+1} \}$.

Assume first that $n+1 \in S$. Since $S\setminus\{n+1\}$ can repair the $i_1$th node in the original system we have a circuit in the original code consisting of $i_1$ and some $t(\leq s)$ elements of $S\setminus\{n+1\}$. Without loss of generality we may assume that these elements are $\{j_1,\dots,j_{t} \}$. Hence there exist elements $b_1,\dots,b_t$ such that \[\mathbf{x}_{i_1}=b_1\mathbf{x}_{j_1}+\dots + b_{t}\mathbf{x}_{j_t}.\] Clearly,
\[
\mathbf{y}_{i_1}=b_1\mathbf{y}_{j_1}+\dots + b_{t}\mathbf{y}_{j_t}+a\mathbf{e}
\]
for some $a \in \mathbb{F}_q$ and hence $S$ can repair $i_1$ in the new code.

Assume now that $n+1 \not\in S$. We can write \[\mathbf{x}_{i_1}=b_1\mathbf{x}_{j_1}+\dots +b_{s}\mathbf{x}_{j_s}\] with some elements $b_1,\dots,b_s$. Since $\mathbf{x}_{i_1}\neq \mathbf{0}$ we can assume without loss of generality that $b_1 \neq 0$. We also have elements $c_2,\dots,c_{s+1}$ such that \[\mathbf{x}_{i_1}=c_2\mathbf{x}_{j_2}+\dots + c_{s+1}\mathbf{x}_{j_{s+1}},\] and hence
\begin{equation}
\begin{cases}
&\mathbf{y}_{i_1}=b_1\mathbf{y}_{j_1}+\dots + b_{s}\mathbf{y}_{j_s}+b\mathbf{e} \\
&\mathbf{y}_{i_1}=c_2\mathbf{y}_{j_2}+\dots + c_{s+1}\mathbf{y}_{j_{s+1}}+c\mathbf{e},
\end{cases}
\end{equation}
for some $b,c \in \mathbb{F}_q$. This gives that
\[
(c-b)\mathbf{e}=b_1\mathbf{y}_{j_1}+(b_2-c_2)\mathbf{y}_{j_2}+\dots + (b_{s}-c_{s})\mathbf{y}_{j_{s}} - c_{s+1}\mathbf{y}_{j_{s+1}}.
\]

If $c\neq b$ then
\[
\mathbf{y}_{i_1}= \sum_{i=1}^{s} b_i\mathbf{y}_{j_i} +\frac{b}{c-b} \left( b_1\mathbf{y}_{j_1} - c_{s+1}\mathbf{y}_{j_{s+1}} +\sum_{i=2}^{s}(b_i-c_i)\mathbf{y}_{j_i}\right),
\]
and $S$ can repair $i_1$ in the new code.

Assume now that $c=b$. We have
\[
b_1\mathbf{y}_{j_1}=(c_2-b_2)\mathbf{y}_{j_2}+\dots + (c_{s}-b_{s})\mathbf{y}_{j_{s}} + c_{s+1}\mathbf{y}_{j_{s+1}}.
\]
Choose from the elements $c_h-b_h$ $(h=2,\dots,s)$ and $c_{s+1}$ the nonzero ones, and mark them as $d_1,\dots,d_t$ ($t \leq s$). The corresponding indices of vectors are marked as $h_1,\dots,h_t$. Now
\[
b_1\mathbf{y}_{j_1}=d_1\mathbf{y}_{h_1}+\dots + d_{t}\mathbf{y}_{h_{t}}.
\]
Without loss of generality we may assume that $\{ \mathbf{y}_{h_1},\dots,\mathbf{y}_{h_{u}} \}$ is a minimal subset of $\{ \mathbf{y}_{h_1},\dots,\mathbf{y}_{h_{t}} \}$ such that
\[
b_1\mathbf{y}_{j_1}=f_1\mathbf{y}_{h_1}+\dots + f_{u}\mathbf{y}_{h_{u}}.
\]
for some $f_1,\dots,f_u \in \mathbb{F}_q$. Clearly $\{ \mathbf{y}_{h_1},\dots,\mathbf{y}_{h_{u}} \}$ are linearly independent and $f_j \neq 0$ for all $j=1,\dots,u$. In the matrix $G_2$, the indices $j_1,h_1,\dots,h_u$ form a circuit. Hence this cannot be the case in $G_1$, and because
\[
b_1\mathbf{x}_{j_1}=f_1\mathbf{x}_{h_1}+\dots + f_{u}\mathbf{x}_{h_{u}},
\]
we know that $\mathbf{x}_{h_1},\dots,\mathbf{x}_{h_{u}}$ cannot be linearly independent. Without loss of generality we may assume that
\[
f_1\mathbf{x}_{h_1}=g_1\mathbf{x}_{h_2}+\dots + g_{u}\mathbf{x}_{h_{u}}
\]
for some elements $g_1,\dots,g_u$. Now
\begin{equation}
\begin{cases}
&f_1\mathbf{x}_{h_1}-g_1\mathbf{x}_{h_2}-\dots - g_{u}\mathbf{x}_{h_{u}}=\mathbf{0} \\
&f_1\mathbf{y}_{h_1}-g_1\mathbf{y}_{h_2}-\dots - g_{u}\mathbf{y}_{h_{u}} \neq \mathbf{0},
\end{cases}
\end{equation}
since $f_1 \neq 0$. This gives that
\[
f_1\mathbf{y}_{h_1}-g_1\mathbf{y}_{h_2}-\dots - g_{u}\mathbf{y}_{h_{u}} = \epsilon\mathbf{e},
\]
for some $\epsilon \neq 0$. Hence
\[
\mathbf{y}_{i_1}=b_1\mathbf{y}_{j_1}+\dots + b_{s}\mathbf{y}_{j_s}+\frac{b}{\epsilon}\left( f_1\mathbf{y}_{h_1}-g_1\mathbf{y}_{h_2}-\dots - g_{u}\mathbf{y}_{h_{u}} \right),
\]
proving that $S$ can repair $i_1$th node in the new code.

We will next show that $\{ i_1,\dots,i_{s+\delta-1}, n+1 \}$ is a $(s+1,\delta)$-locality for the $(n+1)$th node in the new system. Let \[S \subseteq \{ i_1,\dots,i_{s+\delta-1}, n+1 \}\setminus\{ n+1 \}\] be a subset with $|S|=s+1$. Write again $S=\{j_1,\dots,j_{s+1} \}$.

Assume first that $i_1 \in S$. We know that $S\setminus \{ i_1 \}$ can repair the $i_1$th node in the original code, and hence there exists a circuit consisting of nodes $i_1$ and some $t$ nodes $h_1,\dots,h_t$ from $S\setminus \{ i_1 \}$. We know that these cannot form a circuit in the new code and hence there exist nonzero elements $b_1,\dots,b_{t+1}$, such that
\[
b_1\mathbf{y}_{i_1}+b_2\mathbf{y}_{h_1}+\dots+b_{t+1}\mathbf{y}_{t+1}=\epsilon \mathbf{e}
\]
for some $\epsilon \neq 0$, and hence $S$ can repair the $(n+1)$th node.

Assume now that $i_1 \not\in S$. We know that $S\setminus \{ j_{s+1} \}$ can repair $i_1$ in the original code, and similarly as above we have \[\mathbf{x}_{i_1}=b_1\mathbf{x}_{j_1}+\dots b_{s}\mathbf{x}_{j_s},\] for some elements $b_1,\dots,b_s$. Since $\mathbf{x}_{i_1}\neq \mathbf{0}$, we can assume without loss of generality that $b_1 \neq 0$. We have also elements $c_2,\dots,c_{s+1}$ such that \[\mathbf{x}_{i_1}=c_2\mathbf{x}_{j_2}+\dots + c_{s+1}\mathbf{x}_{j_{s+1}}\] and hence again
\begin{equation}
\begin{cases}
&\mathbf{y}_{i_1}=b_1\mathbf{y}_{j_1}+\dots + b_{s}\mathbf{y}_{j_s}+b\mathbf{e} \\
&\mathbf{y}_{i_1}=c_2\mathbf{y}_{j_2}+\dots + c_{s+1}\mathbf{y}_{j_{s+1}}+c\mathbf{e}
\end{cases}
\end{equation}
for some $b,c \in \mathbb{F}_q$. This gives that
\[
(c-b)\mathbf{e}=b_1\mathbf{y}_{j_1}+(b_2-c_2)\mathbf{y}_{j_2}+\dots + (b_{s}-c_{s})\mathbf{y}_{j_{s}} - c_{s+1}\mathbf{y}_{j_{s+1}}.
\]
Again if $c \neq b$ we can repair the $(n+1)$th node so assume that $c=b$. Similarly as above, we can express $\mathbf{e}$ as a linear combination of $y_{j_1},\dots,y_{j_{s+1}}$. Hence $S$ can repair the $(n+1)$th node.
\end{IEEEproof}

The following example illustrates the strength of the above result in the case that $r$ and $k$ are close enough to each other.

\begin{exam}
Let $r \in [\frac{k}{2},k)$ and $C$ be an optimal linear locally repairable code for parameters $(n,k,d,r,\delta)$ over a field $\mathbb{F}_q$ with \[q > (d+r) \binom{n}{\lfloor \frac{n}{2} \rfloor}.\] Because of the optimality we have
\[
d = n-k- \left( \left \lceil \frac{k}{r} \right \rceil -1 \right)\left( \delta - 1 \right) + 1=n-k- \delta +2.
\]

Theorem \ref{Thm:Enlarge} results a locally repairable code for parameters $(n'=n+1,k'=k+1,d'=d,r'=r+1,\delta'=\delta)$. This code is also optimal, as we have
\[\begin{split}
&n'-k'- \left( \left \lceil \frac{k'}{r'} \right \rceil -1 \right)\left( \delta' - 1 \right) + 1\\
=\,& n-k- \left( \left \lceil \frac{k+1}{r+1} \right \rceil -1 \right)\left( \delta - 1 \right) + 1\\
=\,& n-k- \delta + 2\\
=\, & d=d'.
\end{split}\]
Hence the proof of the above theorem gives a procedure to build optimal codes using already known optimal codes in the case that the size of the repair locality is at least half of the code dimension.
\end{exam}

\subsection{Puncturing codes}
Puncturing is a traditional method in classical coding theory. The next theorem shows that this method is useful also in the context of locally repairable codes. Puncturing is used in the field of storage codes at least in \cite{rashmiOptimal,exactjournal}.

\begin{thm}\label{Thm:codePuncturing}
Suppose we have a linear locally repairable code $C$ with all-symbol locality associated to parameters $(n,k,d,r,\delta)$. There exists a linear locally repairable code $C'$ with all-symbol locality associated to parameters \[(n'=n-1,k'=k-1,d' \geq d,r'=r,\delta'=\delta).\]
\end{thm}
\begin{IEEEproof}
Write
\[
C_x=\{\mathbf{y} \in C \mid \mathbf{y}=(x,\mathbf{z}) \text{ where } \mathbf{z} \in \mathbf{F}_{q}^{n-1} \}
\]
for $x \in \mathbf{F}_{q}$.

Clearly each element of $C$ is contained in exactly one of the subsets $C_x$ with $x \in \mathbf{F}_{q}$. Hence there exists~$a \in \mathbf{F}_{q}$ such that
\[
|C_a| \geq \frac{|C|}{q}=q^{k-1}.
\]
It is easy to verify that $|C_0| \geq |C_a| \geq q^{k-1}$. To be precise, we have either $|C_0|=q^{k-1}$ or $|C_0|=q^{k}$.

Define $C'$ to be a code we get by puncturing the first component of $C_0$, \emph{i.e.},
\[
C' =\{\mathbf{z} \in \mathbf{F}_{q}^{n-1} \mid (0,\mathbf{z}) \in C_0 \}.
\]
Clearly $C'$ is a subspace of $\mathbf{F}_{q}^{n-1}$ and its minimum distance $d'$ is at least the same as the minimum distance of $C$, \emph{i.e.} $d' \geq d$.

The dimension $k'$ of $C'$ is at least $k-1$. If $k'=k$ then just delete $1$ row from the generator matrix. Also, it has all-symbol $(r,\delta)$-locality. Indeed, suppose we need to repair the $j$th node. If the first node from the original system is not in the repair locality, then the repair can be made as in the original code. If the first node is in the repair locality, then we know that $0$ is stored into that node and hence the repair can be made using the other nodes from the original locality.
\end{IEEEproof}

\begin{exam}\label{Exam:InductionPuncture}
Suppose that $C$ is an optimal code. It is associated with parameters $(n,k,d,r,\delta)$ with equality
\[
d = n - k - \left( \left \lceil \frac{k}{r} \right \rceil -1 \right)\left( \delta - 1 \right) + 1.
\]
Let $C'$ be a code formed from $C$ using the method explained in Theorem \ref{Thm:codePuncturing}. Hence it is associated with parameters \[(n'=n-1,k' = k-1,d' \geq d,r' =r,\delta'=\delta).\] This code is optimal if
\[
d = n - k - \left( \left \lceil \frac{k-1}{r} \right \rceil -1 \right)\left( \delta - 1 \right) + 1,
\]
which is true if
\[
\left\lceil\frac{k}{r}\right\rceil=\left\lceil\frac{k-1}{r}\right\rceil,
\]
\emph{i.e.}, if
$r$ does not divide $k-1$.
\end{exam}

Together Theorems \ref{Thm:Enlarge} and \ref{Thm:codePuncturing} give the following corollary.
\begin{corollary}
Let $D_q(n,k,r,\delta)$ denote the largest achievable minimum distance for a linear code of length $n \geq 3$, dimension $k \geq 2$, and all-symbol $(r,\delta \geq 2)$-locality, over a field of size $q$. If \[q > (n-k+r) \binom{n}{\left\lfloor \frac{n}{2} \right\rfloor},\] then
\[
D_q(n,k,r,\delta) \leq D_q(n-1,k-1,r,\delta) \leq D_q(n,k,r+1,\delta).
\]
\end{corollary}
\begin{IEEEproof}
The first inequality is proved in Theorem \ref{Thm:codePuncturing}. If $k-1>r$ then the second inequality is proved in Theorem \ref{Thm:Enlarge} since
\[\begin{split}
D_q(n-1,k-1,r,\delta) & \leq n-k-\left( \left \lceil \frac{k-1}{r} \right \rceil -1 \right)\left( \delta - 1 \right) + 1 \\
& \leq n-k-\delta+2\\
& \leq n-k,
\end{split}\]
and hence
\[\begin{split}
q & > (n-k+r) \binom{n}{\left\lfloor \frac{n}{2} \right\rfloor} \\
& \geq (D_q(n-1,k-1,r,\delta)+r) \binom{n}{\left\lfloor \frac{n}{2} \right\rfloor}.
\end{split}\]

If $k-1 \leq r$ then optimal LRCs associated to parameters $(n-1,k-1,r,\delta)$ or $(n,k,r+1,\delta)$ correspond to maximum distance separable codes. Those can be generated by using Cauchy matrices which are known to exist since by assumption $q \geq n+k$.
\end{IEEEproof}

\section{Code construction}\label{Sec:minDistance}

\subsection{Construction}\label{Subsec:Construction}

In this subsection we will give a construction for linear locally repairable codes with all-symbol $(r,\delta)$-locality over a field $\mathbb{F}_q$ with \[q>(r\delta)^{r4^r}\binom{n+(r\delta)^{(r-1)4^r}}{k-1}\] when given parameters $(n,k,r,\delta)$ such that \[n-\left\lceil\frac{n}{r+\delta-1}\right\rceil(\delta-1) \geq k.\] We also assume that $k<n$ and \[n \not\equiv 1,2,\dots,\delta-1 \mod r+\delta-1.\] Write \[n=a(r+\delta-1)-b,\] with $0 \leq b < r$.

We will construct a generator matrix for a linear code under the above assumptions. The minimum distance of the constructed code is studied in Subsection \ref{Subsec:Analysis}. The field used in the construction is huge and we have not attempted to minimize its size, since the main use for this construction is in the proof of Theorem \ref{Thm:randomCode}, where the field size is assumed to approach infinity. However, we do want to present the construction in deterministic form.

First we will build $a=\left\lceil\frac{n}{r+\delta-1}\right\rceil$ sets \[S_1, S_2, \dots, S_a\subseteq\mathbb{F}_q^k,\] with \[|S_i|=r+\delta-1\mbox{ for }1\leq i<a,\] and \[|S_a|=r+\delta-1-b.\] Write
\[
M=(I_r | B_{r \times (\delta-1)})=
\left( \begin{matrix}
   a_{1,1} & \hdots & a_{1,r+\delta-1} \\
   \vdots  & \ddots & \vdots \\
   a_{r,1} & \hdots & a_{r,r+\delta-1}
  \end{matrix} \right)
\]
where $I_r$ is an identity matrix of size $r$ and $B_{r \times (\delta-1)}$ is an $r~\times~(\delta-1)$ matrix all of whose square submatrices are invertible. Do not confuse the entries $a_{i,j}$ with the number $a=\left\lceil\frac{n}{r+\delta-1}\right\rceil$. We also write
\[
M_j=
\left( \begin{matrix}
   a_{1,1} & \hdots & a_{1,j} & a_{1,r+1} & \hdots & a_{1,r+\delta-1} \\
   \vdots  & \ddots & \vdots  & \vdots    & \ddots & \vdots \\
   a_{j,1} & \hdots & a_{j,j} & a_{j,r+1} & \hdots & a_{j,r+\delta-1}
  \end{matrix} \right)=(I_j | B_j).
\]
Define further
\[
U_0 = \left\{ a_{i_1,i_2} \mid 1 \leq i_1 \leq r \text{ and } 1 \leq i_2 \leq r+\delta-1 \right\}
\]
and
\[
U_{m+1} = \left\{ x - \frac{yz}{w} \mid x,y,z,w \in U_m \text{ and } w \neq 0 \right\}
\]
for $m=0,\dots,r-1$. Notice that $U_m \subseteq U_{m+1}$ if $r \geq 2$. We have \[|U_0| \leq r\delta\,, |U_{m+1}| \leq |U_m|^4\mbox{ and }|U_r| \leq (r\delta)^{4^r}.\]

Next, choose any $r$ linearly independent vectors $\mathbf{g}_{1,1},\dots,\mathbf{g}_{1,r} \in \mathbb{F}_{q}^{k}$. Let
\[
\mathbf{s}_{1,r+j}=\sum_{l=1}^{r}a_{l,r+j}\mathbf{g}_{1,l}
\]
for $j=1,\dots,\delta-1$. These $r+\delta-1$ vectors form the set $S_1$. Notice that these vectors correspond the columns of matrix
\[
(\mathbf{g}_{1,1}|\dots|\mathbf{g}_{1,r}) M = (\mathbf{g}_{1,1}|\dots|\mathbf{g}_{1,r}|\mathbf{s}_{1,r+1}|\dots|\mathbf{s}_{1,r+\delta-1}).
\]
This set has the property that any $r$ vectors in it are linearly independent.

Let $1< i \leq a$. We will construct sets $S_1, S_2, \dots, S_{i-1}$ such that any $k$ vectors from $\bigcup_{j=1}^{i-1}S_j$, at most $r$ of which are from the same $S_j$, are linearly independent. The construction will be recursive over $i$, and the set $S_1$ will be as defined above.

Let $\mathbf{g}_{i,1}$ be any vector such that when taking at most $k~-1$ vectors from the already built sets $S_1, S_2, \dots, S_{i-1}$, with at most $r$ vectors from each set, then $\mathbf{g}_{i,1}$ and these $k-1$ other vectors are linearly independent. This is possible since \[\binom{n}{k-1}q^{k-1}<q^k.\]

Write \[\mathbf{s}_{i,r+m}^{(h)} = \sum_{l=1}^{h} a_{l,r+m}\mathbf{g}_{i,l}\] for $m=1,\dots,\delta-1$ and $h=1,\dots,r$, and to shorten the notation, write $\mathbf{s}_{i,r+m}=\mathbf{s}_{i,r+m}^{(r)}$ for $m=1,\dots,\delta-1$, \emph{i.e.},
\[
(\mathbf{g}_{i,1}|\dots|\mathbf{g}_{i,h}|\mathbf{s}^{(h)}_{i,r+1}|\dots|\mathbf{s}^{(h)}_{i,r+\delta-1}) = (\mathbf{g}_{i,1}|\dots|\mathbf{g}_{i,h}) M_h.
\]
Define also
\[
V_j= \{ u_1\mathbf{g}_{i,1} + \dots + u_j\mathbf{g}_{i,j} \mid u_h \in U_{r} \text{ and } u_j \neq 0 \}
\]
and
\[
W_j= \{ u_1\mathbf{g}_{i,1} + \dots + u_j\mathbf{g}_{i,j} \mid u_h \in U_{r} \}.
\]
Notice that $|V_j| \leq |W_j| \leq |U_{r}|^{j} \leq (r\delta)^{j4^r}$.

Suppose we have $j-1$ vectors $\mathbf{g}_{i,1},\dots,\mathbf{g}_{i,j-1}$ such that the following two properties hold:
\begin{enumerate}
\item Any subset \[I\subseteq \bigcup_{t=1}^{i-1} S_t \cup\{\mathbf{g}_{i,1},\dots,\mathbf{g}_{i,j-1},\mathbf{s}_{i,r+1}^{(j-1)},\dots,\mathbf{s}_{i,r+\delta-1}^{(j-1)} \},\] with \[|I|\leq k\,, |I\cap S_t|\leq r\mbox{ for }1\leq t\leq i-1\] and \[|I\cap\{\mathbf{g}_{i,1},\dots,\mathbf{g}_{i,j-1},\mathbf{s}_{i,r+1}^{(j-1)},\dots,\mathbf{s}_{i,r+\delta-1}^{(j-1)} \}|\leq j-1,\] is linearly independent.
\item For any $1\leq l <j$ and for any subset \[I\subseteq  \bigcup_{t=1}^{i-1} S_t \cup W_{l-1},\] with \[|I|\leq k-1\,, |I\cap S_t|\leq r\mbox{ for }1\leq t\leq i-1\] and \[|I\cap W_{l-1}|\leq l-1,\] none of the vectors in $V_{l}$ lies in the linear hull of $I$.
\end{enumerate}
Notice that the properties (1) and (2) are true for $j=2$. Now, the basis for the induction is ready.

Let $\mathbf{g}_{i,j}$ be any vector such that property (2) holds also for $j=l$. This is possible because there are at most $\binom{n+(r\delta)^{(j-1)4^r}}{k-1}$ different possibilities to choose, each of the options span a subspace with $q^{k-1}$ vectors, and since $q$ is large we have \[(r\delta)^{j4^r}\binom{n+(r\delta)^{(j-1)4^r}}{k-1}q^{k-1}<q^k.\] Notice that $u\mathbf{g}_{i,j}+\mathbf{v} \in V$ (where $V$ is some subspace) if and only if $u\mathbf{g}_{i,j} \in -\mathbf{v}+V$.

To prove the induction step we have to prove that property (1) still holds when replacing $j-1$ by $j$.
Let $1 \leq l \leq j$, $\mathbf{v}$ be a linear combination of at most $k-l$ vectors from the sets $S_1, S_2, \dots, S_{i-1}$ with at most $r$ vectors from each set. We will assume the contrary: We have coefficients \[s_{1},\dots,s_{l} \in \mathbf{F}_q \setminus \{ 0 \},\] such that
\[
\mathbf{v} + \sum_{m=1}^{l}s_m\sum_{h=1}^{j}a_{h,f_m}\mathbf{g}_{i,h} = \mathbf{0},
\]
with $f_1 \leq \dots \leq f_l$ and \[f_m \not\in \{ j+1,j+2,\dots,r \}\mbox{ for }m=1,\dots,l.\]

Write
\[
\sum_{m=1}^{l}s_m\sum_{h=1}^{j}a_{h,f_m}\mathbf{g}_{i,h} = \sum_{h=1}^{j}b_h \mathbf{g}_{i,h},
\]
\emph{i.e.},
\[
\begin{pmatrix}
    b_1 \\
    \vdots \\
    b_j
\end{pmatrix}
=\begin{pmatrix}
    a_{1,f_1} & \hdots & a_{1,f_l} \\
    \vdots    & \ddots & \vdots \\
    a_{j,f_1} & \hdots & a_{j,f_l}
\end{pmatrix}
\begin{pmatrix}
    s_1 \\
    \vdots \\
    s_l
\end{pmatrix}.
\]
Again, do not confuse the entries $b_i$ with the number $b=a(r+\delta-1)-n$. Without loss of generality we may assume that $a_{j,f_l} \neq 0$, since otherwise we would also have \[a_{j,f_1}=\dots=a_{j,f_l}=0.\]

Let $t$ be the smallest non-negative integer such that $b_{j-t}~\neq0$. Such $t$ exists since the rank of $(a_{h,f_i})_{j \times l}$ is $l$ and
\[
\begin{pmatrix}
    s_1 \\
    \vdots \\
    s_l
\end{pmatrix} \neq \mathbf{0}.
\]

Hence we have
\[
\begin{pmatrix}
    b_1 \\
    \vdots \\
    b_{j-t} \\
    0 \\
    \vdots \\
    0
\end{pmatrix}
=\begin{pmatrix}
    c_{1,f_1}^{(1)} & \hdots & c_{1,f_{l-1}}^{(1)} & 0 \\
    \vdots & \ddots & \vdots  & \vdots \\
    c_{j-t,f_1}^{(1)} & \hdots & c_{j-t,f_{l-1}}^{(1)} & 0 \\
    c_{j-t+1,f_1}^{(1)} & \hdots & c_{j-t+1,f_{l-1}}^{(1)} & 0 \\
    \vdots & \ddots & \vdots & \vdots \\
    c_{j-1,f_1}^{(1)} & \hdots & c_{j-1,f_{l-1}}^{(1)} & 0 \\
    a_{j,f_1} & \hdots & a_{j,f_{l-1}} & a_{j,f_l}
\end{pmatrix}
\begin{pmatrix}
    s_1 \\
    \vdots \\
    s_l
\end{pmatrix},
\]
where \[c_{h,f_i}^{(1)}=a_{h,f_i}-\frac{a_{h,f_l}a_{j,f_i}}{a_{j,f_l}} \in U_1.\]

This gives\small
\[\begin{split}
\begin{pmatrix}
    b_1 \\
    \vdots \\
    b_{j-t} \\
    0 \\
    \vdots \\
    0
\end{pmatrix}
& =\begin{pmatrix}
    c_{1,f_1}^{(1)} & \hdots & c_{1,f_{l-1}}^{(1)} \\
    \vdots & \ddots & \vdots  \\
    c_{j-t,f_1}^{(1)} & \hdots & c_{j-t,f_{l-1}}^{(1)} \\
    c_{j-t+1,f_1}^{(1)} & \hdots & c_{j-t+1,f_{l-1}}^{(1)} \\
    \vdots & \ddots & \vdots  \\
    c_{j-1,f_1}^{(1)} & \hdots & c_{j-1,f_{l-1}}^{(1)}
\end{pmatrix}
\begin{pmatrix}
    s_1 \\
    \vdots \\
    s_{l-1}
\end{pmatrix}\\
&=\begin{pmatrix}
    c_{1,f_1}^{(2)} & \hdots & c_{1,f_{l-2}}^{(2)} & 0 \\
    \vdots & \ddots & \vdots & \vdots  \\
    c_{j-t,f_1}^{(2)} & \hdots & c_{j-t,f_{l-2}}^{(2)} & 0 \\
    c_{j-t+1,f_1}^{(2)} & \hdots & c_{j-t+1,f_{l-2}}^{(2)} & 0 \\
    \vdots & \ddots & \vdots & \vdots  \\
    c_{j-2,f_1}^{(2)} & \hdots & c_{j-2,f_{l-2}}^{(2)} & 0 \\
    c_{j-1,f_1}^{(1)} & \hdots & c_{j-1,f_{l-2}}^{(1)} & c_{j-1,f_{l-1}}^{(1)}
\end{pmatrix}
\begin{pmatrix}
    s_1 \\
    \vdots \\
    s_{l-1}
\end{pmatrix}.
\end{split}\]\normalsize
Recursively letting \[c_{h,f_i}^{(v)}=c_{h,f_i}^{(v-1)}-\frac{c_{h,f_{l-v+1}}^{(v-1)}c_{j-v+1,f_{i}}^{(v-1)}}{c_{j-v+1,f_{l-v+1}}^{(v-1)}} \in U_v,\]  for $2 \leq v \leq t+1$, and deleting zero columns, we get
\[
\begin{pmatrix}
    b_1 \\
    \vdots \\
    b_{j-t} \\
\end{pmatrix}
=\begin{pmatrix}
    c_{1,f_1}^{(t)} & \hdots & c_{1,f_{l-t}}^{(t)} \\
    \vdots & \ddots & \vdots \\
    c_{j-t,f_{1}}^{(t)} & \hdots & c_{j-t,f_{l-t}}^{(t)} \\
\end{pmatrix}
\begin{pmatrix}
    s_1 \\
    \vdots \\
    s_{l-t},
\end{pmatrix}
\]
if $l-t \geq 1$, and
\[
\begin{pmatrix}
    b_1 \\
    \vdots \\
    b_{j-t} \\
\end{pmatrix}
=\begin{pmatrix}
    c_{1,f_1}^{(t)} \\
    \vdots \\
    c_{j-t,f_{1}}^{(t)} \\
\end{pmatrix}
\begin{pmatrix}
    s_1
\end{pmatrix},
\]if $l-t < 1$. To avoid heavy notion, we will assume that $l-t \geq 1$ from now on. The case $l-t < 1$ would be treated similarly.

The induction step goes through all the way since the smallest non-invertible square matrix in the lower right corner of
\[
\begin{pmatrix}
    a_{1,f_1} & \hdots & a_{1,f_l} \\
    \vdots & \ddots & \vdots \\
    a_{j,f_1} & \hdots & a_{j,f_l}
\end{pmatrix}
\]
has side length at least $t+2$, if it exist, whence \[c_{j-v+1,f_{l-v+1}}^{(v-1)} \neq 0\mbox{ for }v=1,\dots,t+1.\] The proof of this is postponed to Lemma \ref{Lemma:noninvertible}.

Hence we have
\begin{equation}
\begin{split}
\begin{pmatrix}
    b_1 \\
    \vdots \\
    b_{j-t} \\
\end{pmatrix}
= & \begin{pmatrix}
    c_{1,f_1}^{(t)}-\frac{c_{j-t,f_{1}}^{(t)}c_{1,f_{l-t}}^{(t)}}{c_{j-t,f_{l-t}}^{(t)}} & \hdots & c_{1,f_{l-t-1}}^{(t)}-\frac{c_{j-t,f_{l-t-1}}^{(t)}c_{1,f_{l-t}}^{(t)}}{c_{j-t,f_{l-t}}^{(t)}} & c_{1,f_{l-t}}^{(t)} \\
    \vdots & \ddots & \vdots & \vdots \\
    c_{j-t-1,f_1}^{(t)}-\frac{c_{j-t,f_{1}}^{(t)}c_{j-t-1,f_{l-t}}^{(t)}}{c_{j-t,f_{l-t}}^{(t)}} & \hdots & c_{j-t-1,f_{l-t-1}}^{(t)}-\frac{c_{j-t,f_{l-t-1}}^{(t)}c_{j-t-1,f_{l-t}}^{(t)}}{c_{j-t,f_{l-t}}^{(t)}} & c_{j-t-1,f_{l-t}}^{(t)} \\
    0 & \hdots & 0 & c_{j-t,f_{l-t}}^{(t)} \\
\end{pmatrix} \\
& \cdot \begin{pmatrix}
    s_1 \\
    \vdots \\
    s_{l-t-1} \\
    \frac{s_{1}c_{j-t,f_{1}}^{(t)} + \dots + s_{l-t}c_{j-t,f_{l-t}}^{(t)}}{c_{j-t,f_{l-t}}^{(t)}}
\end{pmatrix}.
\end{split}
\end{equation}

By our contra assumption we have
\begin{equation}
\begin{split}
\mathbf{0} & = \mathbf{v} + \sum_{m=1}^{l}s_m\sum_{h=1}^{j}a_{h,f_m}\mathbf{g}_{i,h} \\
 &= \mathbf{v} + \sum_{m=1}^{l-t-1}s_m\sum_{h=1}^{j-t-1}\left(c_{h,f_m}^{(t)}-\frac{c_{j-t,f_{m}}^{(t)}c_{h,f_{l-t}}^{(t)}}{c_{j-t,f_{l-t}}^{(t)}}\right)\mathbf{g}_{i,h} + \frac{s_1c_{j-t,f_{1}}^{(t)}+\dots+s_{l-t}c_{j-t,f_{l-t}}^{(t)}}{c_{j-t,f_{l-t}}^{(t)}}\sum_{h=1}^{j-t}c_{h,f_{l-t}}^{(t)}\mathbf{g}_{i,h} \\
 &= \mathbf{v} + \sum_{m=1}^{l-t-1}s_m\sum_{h=1}^{j-t-1}c_{h,f_m}^{(t+1)}\mathbf{g}_{i,h} + \frac{s_1c_{j-t,f_{1}}^{(t)}+\dots+s_{l-t}c_{j-t,f_{l-t}}^{(t)}}{c_{j-t,f_{l-t}}^{(t)}}\sum_{h=1}^{j-t}c_{h,f_{l-t}}^{(t)}\mathbf{g}_{i,h}.
\end{split}
\end{equation}
But this cannot be true, since $(l-t-1)+1 \leq j-t$ and \[\sum_{h=1}^{j-t}c_{h,f_{l-t}}^{(t)}\mathbf{g}_{i,h} \in V_{j-t}\] is chosen such that it does not belong to the subspace spanned by \[\left\{\mathbf{v}, \sum_{h=1}^{j-t-1}c_{h,f_1}^{(t+1)}\mathbf{g}_{i,h},\dots, \sum_{h=1}^{j-t-1}c_{h,f_{l-t-1}}^{(t+1)}\mathbf{g}_{i,h}\right\},\] and we have \[\frac{s_1c_{j-t,f_{1}}^{(t)}+\dots+s_{l-t}c_{j-t,f_{l-t}}^{(t)}}{c_{j-t,f_{l-t}}^{(t)}} \neq 0\] since $b_{j-t} \neq 0$.

Remember that we wrote $n=a(r+\delta-1)-b$ with $0\leq b<r$. Now, we have sets \[S_i=\{\mathbf{g}_{i,1},\dots,\mathbf{g}_{i,r},\mathbf{s}_{i,r+1},\dots,\mathbf{s}_{i,r+\delta-1}\}\] for $i=1,\dots,a-1$, and\[S_a=\{\mathbf{g}_{a,1},\dots,\mathbf{g}_{a,r-b},\mathbf{s}_{i,r+1}^{(r-b)},\dots,\mathbf{s}_{i,r+\delta-1}^{(r-b)}\}.\] The matrix $\mathbf{G}$ is a matrix with vectors from the sets $S_1,S_2,\dots,S_a$ as its column vectors, \emph{i.e.},
\[
\mathbf{G}=\left(G_1|G_2|\dots|G_a\right)
\]
where
\[
G_j=\left(\mathbf{g}_{j,1}|\dots|\mathbf{g}_{j,r}|\mathbf{s}_{i,r+1}|\dots|\mathbf{s}_{i,r+\delta-1}\right)
\]
for $i=1,\dots,a-1$, and
\[
G_a=\left(\mathbf{g}_{a,1}|\dots|\mathbf{g}_{a,r-b}|\mathbf{s}_{i,r+1}^{(r-b)}|\dots|\mathbf{s}_{i,r+\delta-1}^{(r-b)}\right).
\]

To be a generator matrix for a code of dimension $k$, the rank of $\mathbf{G}$ has to be $k$. By the construction the rank is $k$ if and only if
$n-a(\delta-1) \geq k$, which is what we assumed.

\begin{lemma}\label{Lemma:noninvertible}
The smallest non-invertible square matrix in the lower right corner of
\[
\begin{pmatrix}
    a_{1,f_1} & \hdots & a_{1,f_l} \\
    \vdots    & \ddots & \vdots \\
    a_{j,f_1} & \hdots & a_{j,f_l}
\end{pmatrix}
\]
has side length at least $t+2$, if it exists.
\end{lemma}
\begin{IEEEproof}
Suppose that matrices in the lower right corner with side length at most $N$ are invertible, and that $N$ is maximal with respect to this property. The value $N$ is well-defined and positive since the square matrix with side length $1$ is invertible.

Assume for a contradiction that $N \leq t$ and write
\[
C=\begin{pmatrix}
    a_{j-N+1,f_1} & \hdots & a_{j-N+1,f_{l-N}} \\
    \vdots    & \ddots & \vdots \\
    a_{j,f_1} & \hdots & a_{j,f_{l-N}}
\end{pmatrix}.
\]

Assume first that $C$ is a zero matrix. Now
\[\begin{split}
\mathbf{0}
&=\begin{pmatrix}
    a_{j-N+1,f_{1}} & \hdots & a_{j-N+1,f_{l}} \\
    \vdots & \ddots & \vdots \\
    a_{j,f_{1}} & \hdots & a_{j,f_{l}}
\end{pmatrix}
\begin{pmatrix}
    s_{1} \\
    \vdots \\
    s_{l}
\end{pmatrix}\\
&=\begin{pmatrix}
    a_{j-N+1,f_{l-N+1}} & \hdots & a_{j-N+1,f_{l}} \\
    \vdots          & \ddots & \vdots \\
    a_{j,f_{l-N+1}} & \hdots & a_{j,f_{l}}
\end{pmatrix}
\begin{pmatrix}
    s_{l-N+1} \\
    \vdots \\
    s_{l}
\end{pmatrix},
\end{split}\]
which is not possible.

Assume then that $C$ is not a zero matrix. Clearly $N$ is greater than or equal to the number of columns in
\[
\begin{pmatrix}
    a_{1,f_1} & \hdots & a_{1,f_l} \\
    \vdots    & \ddots & \vdots \\
    a_{j,f_1} & \hdots & a_{j,f_l}
\end{pmatrix}
\]
that correspond to columns of $B_{r \times (\delta-1)}$. Hence
\[
\begin{pmatrix}
    a_{j-N,f_{l-N}} & \hdots & a_{j-N,f_{l}} \\
    \vdots          & \ddots & \vdots \\
    a_{j,f_{l-N}} & \hdots & a_{j,f_{l}}
\end{pmatrix}
=(\mathbf{e}_1|\mathbf{e}_2|\dots|\mathbf{e}_{\epsilon}|B')
\]
where each $\mathbf{e}_i$ has one $1$ and the other elements are zeros, these $1$s are in different rows, and all the square submatrices of $B'$ are invertible. Hence this $(N+1) \times (N+1)$ matrix is also invertible against assumption. This proves that $N \geq t+1$.
\end{IEEEproof}

\begin{remark}
Note that the estimates for $q$ are very rough in the construction. This is because we are mainly interested in the randomized case in which $q \rightarrow \infty$. The randomized version of the construction is studied in Section~\ref{Sec:Random}.
\end{remark}

\begin{remark}
Note that in the above construction we could have chosen different matrices $M=(I_r | B_{r \times (\delta-1)})$ for each $G_j$. Also, the sets $S_j$ do not have to be of the given size. We only need to assume that
\[
\sum_{j=1}^{a} |S_j| =n,
\]
and that \[\delta \leq |S_j| \leq r +\delta-1.\] Then the corresponding matrix is of type \[(I_{|S_j|-\delta+1} | B_{(|S_j|-\delta+1) \times (\delta-1)}).\] By choosing the sets in this way we get rid of the requirement that \[n \not\equiv 1,2,\dots,\delta-1 \mod r+\delta-1.\]
\end{remark}

\subsection{The minimum distance of the constructed code}\label{Subsec:Analysis}
Next we will calculate the minimum distance of the constructed code, with the assumption that the sets $S_j$ are of size $s_j$ ($j=1,\dots,A$), respectively. Assume also without loss of generality that $s_1 \leq \dots \leq s_A$. Write
\[
\mathbf{G}=\left(E_1|F_1|E_2|F_2|\dots|E_A|F_A\right),
\]
where \[E_j=\left( \mathbf{g}_{j,1}|\dots|\mathbf{g}_{j,s_j-\delta+1} \right)\] and \[F_j=\left( \mathbf{s}_{j,r+1},\dots,\mathbf{s}_{j,r+\delta-1} \right)\] for $j=1,\dots,A$.

Let $e_1,\dots,e_k \in \mathbb{F}_q$ be such elements that $e_l \neq 0$ for some $l=1,\dots,k$, and
\[
(e_1,\dots,e_k)\mathbf{G}
\]
is of minimal weight. By changing columns between $E_j$s and $F_j$s, we may assume that the weight of
\[
(e_1,\dots,e_k)\left(E_1|E_2|\dots|E_A\right)
\]
is minimal, that is, it has the biggest possible amount $k~-1$ of zeros. Indeed, the matrix \[\left(E_1|E_2|\dots|E_A\right)\] generates a maximum distance separable code.

Suppose that
\[
(e_1,\dots,e_k)F_j
\]
has a zero, \emph{i.e.}, its weight is not $\delta-1$. If \[(e_1,\dots,e_k)E_j \neq \mathbf{0},\] then by changing columns between $E_j$ and $F_j$ we would get one more zero into \[(e_1,\dots,e_k)\left(E_1|E_2|\dots|E_A\right),\] which is not possible. Hence the number of zeros in \[(e_1,\dots,e_k)\left(F_1|F_2|\dots|F_A\right)\] is at most $z(\delta-1)$ where $z$ is an integer such that
\[
\sum_{j=1}^{z} (s_j-\delta+1) \leq k-1 \text{ and } \sum_{j=1}^{z+1} (s_j-\delta+1) > k-1.
\]
Hence the minimum distance of the code is
\[
n-(k-1)-z(\delta-1).
\]

\begin{exam}
Recall that a code is called almost optimal if its minimum distance is at least $d_{\text{opt}}(n,k,r,\delta)-\delta+1$. Suppose that $n=a(r+\delta-1)$, and choose that $s_j-\delta+1=r$ for all $j=1,\dots,a$. Then $z=\left\lfloor\frac{k-1}{r}\right\rfloor$, and hence the minimum distance is
\[\begin{split}
&n-(k-1)-\left\lfloor\frac{k-1}{r}\right\rfloor(\delta-1)\\
 =\,& n-k-\left(\left\lceil\frac{k}{r}\right\rceil-1\right)(\delta-1) + 1 \\
 =\,& d_{\text{opt}}(n,k,r,\delta),
\end{split}\]
so the construction is optimal.

Suppose then, that $n=a(r+\delta-1)+b$ with $0 \leq b<r+\delta-1$. If $0 < b < \delta$, then using the above optimal code with $b$ extra copies of other columns in the generator matrix, we get a code with minimum distance \[d_{\text{opt}}(n-b,k,r,\delta)=d_{\text{opt}}(n,k,r,\delta)-b \geq d_{\text{opt}}(n,k,r,\delta)-\delta+1.\]

If $b \geq \delta$, then choose $s_j=r+\delta-1$ for $j=1,\dots,a$ and $s_{a+1}=b$. Now $z=\left\lceil\frac{k-b+\delta-1}{r}\right\rceil$ and hence the minimum distance is
\[
n-k-\left\lceil\frac{k-b+\delta-1}{r}\right\rceil(\delta-1) + 1.
\]
Now,
\[\begin{split}
& d_{\text{opt}}(n,k,r,\delta)-\left(n-k-\left\lceil\frac{k-b+\delta-1}{r}\right\rceil(\delta-1) + 1\right)\\
=\, &(\delta-1)\left(\left\lceil\frac{k-b+\delta-1}{r}\right\rceil-\left\lceil\frac{k}{r}\right\rceil +1\right) \\
\leq\, &\delta-1,
\end{split}\]
and hence the code is again at least almost optimal.
\end{exam}

\section{Random matrices as generator matrices for locally repairable codes}\label{Sec:Random}

\subsection{The structure of the codes}
We will study linear codes, where the nodes are divided into non-overlapping sets $S_1,S_2,\dots,S_a$, such that any node $x \in S_j$ can be repaired by any $|S_j \setminus \{x\}|-(\delta-2)=|S_j|-\delta+1$ nodes from $S_j$. We also require that $|S_j| \leq r+\delta-1$ and to guarantee the all-symbol repairing property, that $\bigcup_{j=1}^{a} S_j =\{1,\dots,n\}$. Suppose we have a $k$-dimensional linear code, and a repair set $S_1$ is formed by the nodes, say, $1,2,\dots,s$ ($\delta \leq s \leq r+\delta-1$) corresponding to columns in the generator matrix. Denote by $G$ the $k \times s$ matrix defined by these columns, and write $t=s-\delta+1$. It is natural to require that $G$ is of maximal rank, \emph{i.e.}, that the rank of $G$ is $t$.

By the locality assumption, any $t$ columns can repair any other column, \emph{i.e.}, any $t$ columns span the same subspace as all the $s$ columns. So we have
\[
G=(\mathbf{x}_1|\dots|\mathbf{x}_t|\mathbf{y}_{1}|\dots|\mathbf{y}_{\delta-1}),
\]
where each $\mathbf{y}_j$ can be represented as a linear combination of $\mathbf{x}_1,\dots,\mathbf{x}_t$, and $\mathbf{x}_1,\dots,\mathbf{x}_t$ are linearly independent. This gives that
\[
G=(\mathbf{x}_1|\dots|\mathbf{x}_t)(I_t|B)
\]
where $I_t$ is an identity matrix of size $t$ and $B$ is $t \times (\delta-1)$ matrix.

Let $G'$ consist of some $t$ columns of $G$, and let $C$ consist of the corresponding columns of $(I_t|B)$. It is easy to verify that
\[
G'=(\mathbf{x}_1|\dots|\mathbf{x}_t)C,
\]
and hence
\[
\rank(C)=\rank((\mathbf{x}_1|\dots|\mathbf{x}_t)C)=\rank(G')=t.
\]

Consider a submatrix of $B$ consisting of rows $i_1,\dots,i_l$ and columns $j_1,\dots,j_l$. It is easy to check that this submatrix is invertible if and only if a submatrix corresponding to the columns $\{1,\dots,t\} \setminus \{i_1,\dots,i_l\}$ and $\{t+j_1,\dots,t+j_l\}$ of $(I_t|B)$ is invertible. This is invertible, since the rank of the submatrix of $G$ consisting of the same columns is $t$. Hence any square submatrix of $B$ is invertible.

Suppose the matrices $(I_{t_1}|B_1),\dots,(I_{t_A}|B_a)$ are of this form. It is natural to study codes with generator matrix of the form
\[
\left(G_1 | \dots | G_A \right),
\]
where
\[
G_j=(\mathbf{x}_{1,1}|\dots|\mathbf{x}_{1,t_1})(I_{t_1}|B_1)
\]
for $j=1,\dots,a$. The following natural question arises: How should we choose the vectors \[\mathbf{x}_{1,1},\dots,\mathbf{x}_{1,t_1},\dots,\mathbf{x}_{a,1},\dots,\mathbf{x}_{a,t_a}\] such that the given code has the biggest possible minimum distance? The next subsection tries to answer this in the case that we are dealing with large fields.

Notice also that since the rank of a generator matrix is $k$, we have
\begin{equation}
\begin{split}
k & \leq \sum_{i=1}^{a}\rank\left( (\mathbf{x}_{i,1}|\dots|\mathbf{x}_{i,t_i})(I_{t_i}|B_i) \right)\\
& \leq t_1 + \dots + t_a,
\end{split}
\end{equation}
and hence
\[
k \leq n -A(\delta-1) \leq n -\left\lceil\frac{n}{r+\delta-1}\right\rceil(\delta-1).
\]

\subsection{Random codes}\label{Subsec:RandomResult}
In this subsection, we study locally repairable codes generated by random matrices with a few extra columns. These extra columns consist of linear combinations of the randomly chosen columns, guaranteeing the repair property. It is shown that this kind of code has a good minimum distance with probability approaching $1$ as the field size $q$ approaches infinity.

\begin{thm}\label{Thm:randomCode}
Given parameters $(n,k,r,\delta)$ and $a>0$ with \[r<k\leq n-a(\delta-1),\] and positive integers $s_1 \leq s_2 \leq \dots \leq s_a$ such that \[n=\sum_{j=1}^{a}s_j\] and \[\delta \leq |s_j| \leq r+\delta-1\] for $j=1,\dots,a$. Assume that we have \[(s_j-\delta+1)\times (\delta-1)\mbox{-matrices}\] $B_1,B_2,\dots,B_a$, all of whose square submatrices are invertible. Also, let $x_{i,j}$ be independent and identically distributed uniform random variables over $\mathbb{F}_q$.

Consider matrices $E$, $F$ and $G$ that are defined as follows:
\begin{equation}\label{Eq:matrixE}
E=\begin{pmatrix}
  x_{1,1} & x_{1,2} & \cdots & x_{1,n-a(\delta-1)} \\
  x_{2,1} & x_{2,2} & \cdots & x_{2,n-a(\delta-1)} \\
  \vdots  & \vdots  & \ddots & \vdots    \\
  x_{k,1} & x_{k,2} & \cdots & x_{k,n-a(\delta-1)}
 \end{pmatrix}=(E_1|E_2|\dots|E_a),
\end{equation}
where $E_j$ is a $k \times (s_j-\delta+1)$ matrix for $j=1,\dots,a$,
\[
F=(E_1B_1|E_2B_2|\dots|E_aB_a),
\]
and
\[
G=(E|F).
\]

With probability approaching one as $q \rightarrow \infty$, $G$ is a generator matrix for a $k$-dimensional locally repairable code of length $n$ with all-symbol $(r,\delta)$-locality and minimum distance
\[
d \geq n-k-z(\delta-1)+1,
\]
where $z$ is the unique integer such that
\[
\sum_{j=1}^{z} (s_j-\delta+1) \leq k-1 \text{ and } \sum_{j=1}^{z+1} (s_j-\delta+1) > k-1.
\]
\end{thm}
\begin{IEEEproof}
In the construction of Subsection \ref{Subsec:Construction}, we selected a total of $n$ vectors $\mathbf{g}_{i,j}\in \mathbb{F}_q^k$, that were required not to lie in any of $\binom{n+(r\delta)^{(j-1)4^r}}{k-1}$ prescribed affine hyperplanes. Clearly, there are at most
\[
(r\delta)^{r4^r}\binom{n+(r\delta)^{(r-1)4^r}}{k-1}q^{k-1}
\]
vectors that violate this condition. We call a vector that satisfies the condition \emph{good}.

If we choose the vector $\mathbf{g}_{i,j}$ uniformly from $\mathbb{F}_q^k$, the probability that it is good is thus at least \[\frac{q^k-(r\delta)^{r4^r}\binom{n+(r\delta)^{(r-1)4^r}}{k-1}q^{k-1}}{q^k}.\]
The matrix $G$ is a generator matrix of same type (except the order of the columns) as the generator matrix built in the construction of Subsection \ref{Subsec:Construction}, assuming all the selected column vectors are good. Hence the probability that the whole code is locally repairable with all-symbol $(r,\delta)$-locality and minimum distance $d$, is at least
\[
\left(1-\frac{(r\delta)^{r4^r}\binom{n+(r\delta)^{(r-1)4^r}}{k-1}}{q}\right)^n \rightarrow (1-0)^n=1,
\]
as $q \rightarrow \infty$.
\end{IEEEproof}

\section{Optimal Vector-Linear $(n,k,d,r)$-LRCs Over $\mathbb{F}_2^2$} \label{sec:construction}

In this section we will first define quasi-uniform codes and give some basic facts about this class of codes. Then, by using a construction of quasi-uniform codes, we will give three classes of optimal vector-linear LRCs over $\mathbb{F}_2^2$.

\subsection{Quasi-Uniform Codes}

Let $\mathbb{A}_1,\ldots,\mathbb{A}_n$ be nonempty finite sets. A code $C \subseteq \mathbb{A}_1 \times \ldots \times \mathbb{A}_n$ is said to be \emph{quasi-uniform} if the condition that
$$
\lvert \{(c_1,\ldots,c_n) \in C : (c_{i_1},\ldots,c_{i_m}) = \boldsymbol{a} \} \rvert = \frac{\lvert C  \rvert}{\lvert C_X \rvert},
$$
is satisfied by all \[X=\{i_1,\ldots,i_m\} \subseteq \lbrack n \rbrack,\] and all \[\boldsymbol{a}=(a_1,\ldots,a_m) \in C_X.\] Quasi-uniform codes were introduced in \cite{chan13}.

An explicit construction of quasi-uniform codes from groups is given in \cite{thomas13}. This construction can be characterized as follows. Let $G$ be a finite group and let $G_1,\ldots,G_n$ be some (not necessarily distinct) normal subgroups of $G$. Further, let $\mathbb{A}_i$ be isomorphic to the quotient group $G / G_i$ for $i = 1,\ldots,m$. Now, we get a quasi-uniform code $C$ by the following construction,
\begin{equation} \label{eq:construction-group}
C = \{(gG_1,\ldots,gG_n) : g \in G\} \subseteq \mathbb{A}_1 \times \ldots \times \mathbb{A}_n.
\end{equation}
The minimum distance $d$ of $C$ and the size of its projections was given in \cite{thomas13} as follows. For $X \in \lbrack n \rbrack$, let $G_X = \bigcap_{i \in X} G_i$, then
\begin{equation} \label{eq:size-d-quasi}
\begin{array}{rl}
(i) & d = n - \max \{\lvert X \rvert : X \in \lbrack n \rbrack \hbox{, } \lvert G_X \rvert > 1\},\\
(ii) & \lvert C_X \rvert = \frac{\lvert G \rvert}{\vert G_X \rvert}.
\end{array}
\end{equation}
The code $C$ is a subgroup of $\mathbb{A}_1 \times \ldots \times \mathbb{A}_n$.

Note that all linear and vector-linear codes are quasi-uniform. However, there are also quasi-uniform codes which are neither linear nor vector-linear.

\subsection{Constructions of Optimal Vector-Linear LRCs Over $\mathbb{F}_2^2$}

Given a group $A$ and subsets $A_1, \ldots, A_l$ of $A$, let $\langle A_1, \ldots, A_l\rangle$ denote the subgroup of $A$ generated by the elements in $\cup_{i = 1}^l A_i$. Let $\mathbb{Z}_2$ denote the group of integers modulo two.

By using the construction given in (\ref{eq:construction-group}), we will now get three classes of optimal vector-linear LRCs over $\mathbb{F}_2^2$, for small values of $d$ and $r$, and for arbitrary $n$ and $k$ satisfying some congruence restrictions. In the constructions of these three classes of codes we will need the group $A$ and its subgroups $O, A_1$, $A_2$, $A_3$ and $A_4$ defined below.

 Let $O, A_1, \ldots, A_4$ be the following subgroups of $A~=~(\mathbb{Z}_2^2)^3$:
$$
\begin{array}{rll}
O & = & 00 \times 00 \times 00,\\
A_1 & = & 00 \times \mathbb{Z}_2^2 \times \mathbb{Z}_2^2,\\
A_2 & = & \mathbb{Z}_2^2 \times 00 \times \mathbb{Z}_2^2,\\
A_3 & = & \mathbb{Z}_2^2 \times \mathbb{Z}_2^2 \times 00,\\
A_4 & = & \langle111100,110011,010100,010001\rangle.
\end{array}
$$\\

\noindent \emph{Case: The $C_i^1(3,3)$-class of optimal LRC with $(n,k,d,r) = (4i+3,3i+1,3,3)$}\\

Given a positive integer $i$, let $G$ denote the group $(\mathbb{Z}_2^2)^k$ where $k = 3i+1$. For $0 \leq j < i$, let
$$
\begin{array}{l}
G_{4j+1} = A^j \times A_1 \times A^{i-j-1} \times \mathbb{Z}_2^2,\\
G_{4j+2} = A^j \times A_2 \times A^{i-j-1} \times \mathbb{Z}_2^2,\\
G_{4j+3} = A^j \times A_3 \times A^{i-j-1} \times \mathbb{Z}_2^2,\\
G_{4j+4} = A^j \times A_4 \times A^{i-j-1} \times \mathbb{Z}_2^2.
\end{array}
$$
Furthermore, let
$$
\begin{array}{rll}
G_{4i+1} = && (\mathbb{Z}_2^2)^{3i} \times 00,\\
G_{4i+2} = & \langle\{&O^j \times 011000 \times O^{i-j-1}\times 00,\\
	&& O^j \times 110100 \times O^{i-j-1}\times 00,\\
           && O^j \times 110010 \times O^{i-j-1}\times 00,\\
	&& O^j \times 100001 \times O^{i-j-1}\times 00,\\
	&& O^j \times 010000 \times O^{i-j-1}\times 10, \\
	&& O^j \times 110000 \times O^{i-j-1}\times 01 : 0 \leq j < i\} \rangle,\\
G_{4i+3} = & \langle\{&O^j \times 011000 \times O^{i-j-1}\times 00,\\
	&& O^j \times 110100 \times O^{i-j-1}\times 00,\\
           && O^j \times 110010 \times O^{i-j-1}\times 00, \\
	&& O^j \times 100001 \times O^{i-j-1}\times 00,\\
	&& O^j \times 110000 \times O^{i-j-1}\times 10,\\
	&& O^j \times 100000 \times O^{i-j-1}\times 01 : 0 \leq j < i\} \rangle.
\end{array}
$$
Note that all the sets $G_1, \ldots, G_{4i+3} \subseteq G$ defined above are subgroups of $G$. Now, let $C_i^1(3,3)$ denote the quasi-uniform code that we get from $G_1, \ldots, G_{4i+4},G$ in (\ref{eq:construction-group}). Namely,
$$
C_i^1(3,3) = \{gG_1,\ldots,gG_{4i+3} : g \in G\} \subseteq \mathbb{A}_1 \times \ldots \times \mathbb{A}_{4i+3},
$$
where $\mathbb{A}_j \equiv G / G_j$ for $1 \leq j \leq 4i+3$.\\

\noindent \emph{Case: The $C_i^2(3,3)$-class of optimal LRC with $(n,k,d,r) = (4i+4,3i+2,3,3)$}\\

Given a positive integer $i$, let $G$ denote the group $(\mathbb{Z}_2^2)^k$ where $k = 3i+2$. For $0 \leq j < i$, let
$$
\begin{array}{l}
G_{4j+1} = A^j \times A_1 \times A^{i-j-1} \times \mathbb{Z}_2^2 \times \mathbb{Z}_2^2,\\
G_{4j+2} = A^j \times A_2 \times A^{i-j-1} \times \mathbb{Z}_2^2 \times \mathbb{Z}_2^2,\\
G_{4j+3} = A^j \times A_3 \times A^{i-j-1} \times \mathbb{Z}_2^2 \times \mathbb{Z}_2^2,\\
G_{4j+4} = A^j \times A_4 \times A^{i-j-1} \times \mathbb{Z}_2^2 \times \mathbb{Z}_2^2.
\end{array}
$$
Furthermore, let
$$
\begin{array}{rll}
G_{4i+1} = && (\mathbb{Z}_2^2)^{3i} \times 00 \times \mathbb{Z}_2^2 ,\\
G_{4i+2} = && (\mathbb{Z}_2^2)^{3i} \times \mathbb{Z}_2^2 \times 00,\\
G_{4i+3} = & \langle\{&O^j \times 011000 \times O^{i-j-1}\times 0000,\\
	&& O^j \times 110100 \times O^{i-j-1}\times 0000,\\
           && O^j \times 110010 \times O^{i-j-1}\times 0000, \\
	&& O^j \times 100001 \times O^{i-j-1}\times 0000,\\
	&& O^j \times 100000 \times O^{i-j-1}\times 1000,\\
	&& O^j \times 010000 \times O^{i-j-1}\times 0100 : 0 \leq j < i\} \\
					 & \cup& \{O^i \times1011, O^i \times0110\}\rangle,\\
G_{4i+4} = & \langle\{&O^j \times 011000 \times O^{i-j-1}\times 0000,\\
	&& O^j \times 110100 \times O^{i-j-1}\times 0000,\\
           && O^j \times 110010 \times O^{i-j-1}\times 0000,\\
	&& O^j \times 100001 \times O^{i-j-1}\times 0000,\\
	&& O^j \times 100000 \times O^{i-j-1}\times 0010, \\
	&& O^j \times 010000 \times O^{i-j-1}\times 0001 : 0 \leq j < i\} \\
					 & \cup& \{O^i \times1110, O^i \times1001\}\rangle.
\end{array}
$$
Note that all the sets $G_1, \ldots, G_{4i+4} \subseteq G$ defined above are subgroups of $G$. Now, let $C_i^2(3,3)$ denote the quasi-uniform code that we get from $G_1, \ldots, G_{4i+4},G$ in (\ref{eq:construction-group}). Namely,
$$
C_i^2(3,3) = \{gG_1,\ldots,gG_{4i+4} : g \in G\} \subseteq \mathbb{A}_1 \times \ldots \times \mathbb{A}_{4i+4},
$$
where $\mathbb{A}_j \equiv G / G_j$ for $1 \leq j \leq 4i+4$.\\

\noindent \emph{Case: The $C_i^1(4,3)$-class of optimal LRC with $(n,k,d,r) = (4i+4,3i+1,4,3)$}\\

Given a positive integer $i$, let $G$ denote the group $(\mathbb{Z}_2^2)^k$ where $k = 3i+1$. For $0 \leq j < i$, let
$$
\begin{array}{l}
G_{4j+1} = A^j \times A_1 \times A^{i-j-1} \times \mathbb{Z}_2^2,\\
G_{4j+2} = A^j \times A_2 \times A^{i-j-1} \times \mathbb{Z}_2^2,\\
G_{4j+3} = A^j \times A_3 \times A^{i-j-1} \times \mathbb{Z}_2^2,\\
G_{4j+4} = A^j \times A_4 \times A^{i-j-1} \times \mathbb{Z}_2^2.
\end{array}
$$
Furthermore, let $O$, $B_1$, $B_2$, $B_3$, $C_1$, $C_2$ and $C_3$ denote the following subsets of $(\mathbb{Z}_2^2)^3$:
$$
\begin{array}{l}
B_1 =  11 \times \mathbb{Z}_2^2 \times \mathbb{Z}_2^2 \hbox{, } C_1 = 01 \times \mathbb{Z}_2^2 \times \mathbb{Z}_2^2,\\
B_2 = \mathbb{Z}_2^2 \times 11 \times \mathbb{Z}_2^2 \hbox{, } C_2 = \mathbb{Z}_2^2 \times 01 \times \mathbb{Z}_2^2,\\
B_3 = \mathbb{Z}_2^2 \times \mathbb{Z}_2^2 \times 11 \hbox{, } C_3 = \mathbb{Z}_2^2 \times \mathbb{Z}_2^2 \times 01,
\end{array}
$$
and let
$$
\begin{array}{rll}
G_{4i+1} = & \langle\{&O^j \times B_1 \times O^{i-j-1}\times 01,\\
          && O^j \times C_1 \times O^{i-j-1}\times 11 : 0 \leq j < i\}\rangle,\\
G_{4i+2} = & \langle\{&O^j \times B_2 \times O^{i-j-1}\times 01,\\
          && O^j \times C_2 \times O^{i-j-1}\times 11 : 0 \leq j < i\}\rangle,\\
G_{4i+3} = & \langle\{&O^j \times B_3 \times O^{i-j-1}\times 11,\\
          && O^j \times C_3 \times O^{i-j-1}\times 01 : 0 \leq j < i\}\rangle,\\
G_{4i+4} = & \langle\{&O^j \times 111100 \times O^{i-j-1}\times 00,\\
          && O^j \times 110011 \times O^{i-j-1}\times 00,\\
          && O^j \times 110000 \times O^{i-j-1}\times 11,\\
          && O^j \times 010100 \times O^{i-j-1}\times 00,\\
          && O^j \times 010001 \times O^{i-j-1}\times 00,\\
          && O^j \times 010000 \times O^{i-j-1}\times 01: 0 \leq j < i\}\rangle.
\end{array}
$$
Note that all the sets $G_1, \ldots, G_{4i+4} \subseteq G$ defined above are subgroups of $G$. Now, let $C_i^1(4,3)$ denote the quasi-uniform code that we get from $G_1, \ldots, G_{4i+4},G$ in (\ref{eq:construction-group}). Namely,
$$
C_i^1(4,3) = \{gG_1,\ldots,gG_{4i+4} : g \in G\} \subseteq \mathbb{A}_1 \times \ldots \times \mathbb{A}_{4i+4},
$$
where $\mathbb{A}_j \equiv G / G_j$ for $1 \leq j \leq 4i+4$.

\begin{thm}
For $i \geq 1$, the codes $C_i^1(3,3)$ , $C_i^2(3,3)$ and $C_i^1(4,3)$ define optimal vector-linear LRCs over $\mathbb{F}_2^2$ with parameters
$$
\begin{array}{rl}
(i) & (n,k,d,r) = (4i+3,3i+1,3,3) \hbox{ for } C_i^1(3,3),\\
(ii) & (n,k,d,r) = (4i+4,3i+2,3,3) \hbox{ for } C_i^2(3,3),\\
(iii) & (n,k,d,r) = (4i+4,3i+1,4,3) \hbox{ for } C_i^1(4,3).
\end{array}
$$
\end{thm}
\begin{IEEEproof}
We will only prove case $(iii)$; By using similar proof techniques we get case $(i)$ and $(ii)$.

From the construction of $C_i^1(4,3)$ we immediately get that $n = 4i+4$. Further, we observe that $\lvert G \rvert = 4^{3i+1}$ and $\lvert G_j \rvert =  4^{3i}$ for $1 \leq j \leq 4i+4$. It follows, as $g+g=\boldsymbol{0}$ for all $g \in G$ and $|\mathbb{A}_j| = 4$ by (\ref{eq:size-d-quasi}), that $A_j$ can be identified with $\mathbb{Z}_2^2$. Consequently, we now see that our code $C_i^1(4,3)$ can be considered as a subgroup of $(\mathbb{Z}_2^2)^{4i+4}$, or equivalently, as a vector-linear code over $\mathbb{F}_2^2$.

For any integers $a \leq b$, let $\lbrack a, b \rbrack = \{a, a+1, \ldots, b\}$. Moreover, for any finite set $X$ and non-negative integer $a$, let $\binom{X}{a} = \{Y \subseteq X : \lvert Y \rvert = a\}$. Now, we will prove the following facts:
$$
\begin{array}{rl}
(a) & G_{\lbrack 4j+1,4j+4 \rbrack} = G_X \hbox{ for } X \in \binom{\lbrack 4j+1,4j+4 \rbrack}{3}, 0 \leq j \leq i,\\
(b) & G_{\lbrack 1,4i \rbrack} = O^i \times \mathbb{Z}_2^2,\\
(c) & G_X = 00 \times \ldots \times 00 \hbox{, for } X \in \binom{\lbrack n \rbrack}{n-3}.
\end{array}
$$

For (a), we first observe that
$$
\begin{array}{rl}
(a1) & A_1 \cap A_2 = 00 \times 00 \times \mathbb{Z}_2^2,\\
(a2) & A_1 \cap A_3 = 00 \times \mathbb{Z}_2^2 \times 00,\\
(a3) & A_2 \cap A_3 = \mathbb{Z}_2^2 \times 00 \times 00,\\
(a4) & A_1 \cap A_2 \cap A_3 = A_1 \cap A_2 \cap A_4 \\
&= A_1 \cap A_3 \cap A_4 = A_2 \cap A_3 \cap A_4 \\
     &=  00 \times 00 \times 00.
\end{array}
$$
Hence,
\begin{equation} \label{eq:G_j}
\begin{split}
& A^j \times 000000 \times A^{i-j-1} \times \mathbb{Z}_2^2 \\
=\, & G_{\lbrack 4j+1,4j+4 \rbrack} = G_X \hbox{ for } X \in \binom{\lbrack 4j+1,4j+4 \rbrack}{3},
\end{split}\end{equation}
when $0 \leq j \leq i-1$. To prove that statement (a) is satisfied when $j = i$, we first observe that
$$
\begin{array}{rl}
(a5) & G_{4i + 1} = \{\boldsymbol{x} \in (\mathbb{Z}_2^2)^{3i+1} : x_{3i+1} = \sum_j f(x_{3j+1})\},\\
(a6) & G_{4i + 2} = \{\boldsymbol{x} \in (\mathbb{Z}_2^2)^{3i+1} : x_{3i+1} = \sum_j f(x_{3j+2})\},\\
(a7) & G_{4i + 3} = \{\boldsymbol{x} \in (\mathbb{Z}_2^2)^{3i+1} : x_{3i+1} = \sum_j x_{3j+3}\},\\
(a8) & G_{4i + 4} = \{\boldsymbol{x} \in (\mathbb{Z}_2^2)^{3i+1} : x_{3i+1} = \sum_j x_{3j+1} + x_{3j+2} + x_{3j+3}\},\\
(a9) & \sum_j f(x_{3j+1}) = \sum_j f(x_{3j+2}) \\ &\iff \\ &\sum_j x_{3j+1} = \sum_j x_{3j+2},
\end{array}
$$
where every $\boldsymbol{x} \in (\mathbb{Z}_2^2)^{3i+1}$, the sums are taken over $0\leq j\leq i-1$, and $f: \mathbb{Z}_2^2 \rightarrow \mathbb{Z}_2^2$ is the function defined by
$$
f(00) = 00 \hbox{, } f(01) = 11 \hbox{, } f(10) = 10 \hbox{ and } f(11) = 01.
$$
Consequently,
\begin{equation} \label{eq:G_i}
\begin{array}{l}
G_{\lbrack 4i+1,4i+4 \rbrack} = G_X  = \\
\{\boldsymbol{x} \in (\mathbb{Z}_2^2)^{3i+1} : x_{3i+1} = \sum_j f(x_{3j+1}) = \sum_j f(x_{3j+2}) = \sum_j x_{3j+3} \} ,
\end{array}
\end{equation}
for $X \in \binom{\lbrack 4i+1,4i+4 \rbrack}{3}$.

Statement (b) follows from \eqref{eq:G_j}.

For (c) we first observe, by the use of (a5)-(a8), that
$$
G_X \subseteq 00 \times \ldots \times 00 \times \mathbb{Z}_2^2 \quad \Rightarrow \quad G_X = 00 \times \ldots \times 00
$$
as $X \cap [4i+1,4i+4] \neq \emptyset$. Hence, by \eqref{eq:G_j},
$$
| [1,4i] \setminus X | = 0 \hbox{ or } 1 \quad \Rightarrow \quad G_X = 00 \times \ldots \times 00.
$$
Moreover, by (a), if
$$
G_X = 00 \times \ldots \times 00 \hbox{ when } | [1,4i] \setminus X| = 3
$$
then it also holds that
$$
G_X = 00 \times \ldots \times 00 \hbox{ when } | [1,4i] \setminus X| = 2.
$$
Suppose $$[4j+1,4j+4] \setminus X = [4j+1,4j+3]$$ for some $0 \leq j <i$. Then it follows that $$G_X = 00 \times \ldots \times 00$$ by the use of \eqref{eq:G_i} and the fact that
$$
A_4 \cap \{(x_1,x_2,x_3) \in (\mathbb{Z}_2^2)^3 : f(x_1) = f(x_2) = x_3\} = \{000000\}.
$$
Now, suppose \[|[4j+1,4j+4] \setminus X |= 3\] and \[[4j+1,4j+4] \cap X \subseteq [4j+1,4j+3].\] Then the property that $G_X = 00 \times \ldots \times 00$ follows by the use of \eqref{eq:G_i} and the fact that
$$
\{(x_1,x_2,x_3) \in (\mathbb{Z}_2^2)^3 : f(x_1) = f(x_2) = x_3 = 00\} = \{000000\}.
$$

From (c) and (\ref{eq:size-d-quasi}) (ii) we obtain that
$$
\lvert C_i^1(4,3) \rvert = \frac{\lvert G \rvert}{\lvert G_{\lbrack n \rbrack} \rvert} = \frac{\lvert G \rvert}{1} = 4^{3i+1},
$$
and consequently $k = 3i+1$. By the use of (b), (c) and  (\ref{eq:size-d-quasi}) (i) we get that
$$
d = n - 4i = 4.
$$
Observe that the minimum distance of a projection $C_X$ of a code $C \subseteq \mathbb{A}^n$, for some finite alphabet $\mathbb{A}$ and subset $X \subseteq [n]$, is greater than or equal to 2 if and only if \[|C_{X \setminus \{x\}}| = |C_X|\] for every $x \in X$. Hence, as a consequence of (a) and \eqref{eq:size-d-quasi} (ii), $C_i^1(4,3)$ has all-symbol locality $r=3$. This implies that the code is an optimal vector-linear LRC over $\mathbb{F}_2^2$ for $i \geq 1$, since
$$
n - k - \left \lceil \frac{k}{3} \right \rceil + 2 = 4i+4 - (3i + 1) - \left \lceil \frac{3i+1}{3} \right \rceil + 2 = 4 = d.
$$
\end{IEEEproof}

\section{Conclusion}
In this paper we have studied linear locally repairable codes with all-symbol $(r,\delta)$-locality. We have constructed codes with almost optimal minimum distance. Namely, the difference between largest achievable minimum distance of locally repairable codes and the minimum distance of our codes is maximally $\delta-1$. Instead of just giving a construction, it is shown that random matrices augmented by a few columns to guarantee a locality property, asymptotically almost surely (in the field size $q$) generates an almost optimal LRC.

Also, methods to build new codes for different parameters using already existing codes are described. Namely, a method to increase and decrease the code length and dimension are presented.
Constructions of three infinite classes of optimal vector-linear codes over an alphabet of small size, not depending on the size of the code length $n$, are given. This construction is based on quasi-uniform codes.

As a future work it is still left to find the exact expression of the largest achievable minimum distance of the linear locally repairable code with all-symbol $(r,\delta)$-locality when given the length $n$ and the dimension $k$.
In order to find more general classes of optimal LRCs over alphabets of small sizes, further studies of vector-linear LRCs based on quasi-uniform codes are of interest.




\ifCLASSOPTIONcaptionsoff
  \newpage
\fi



%

\end{document}